\documentclass[journal]{IEEEtran}

\usepackage{cite}
\usepackage{amsmath,amssymb,amsfonts}
\usepackage{algorithm}
\usepackage{algorithmic}
\usepackage{graphicx}
\usepackage{booktabs}
\usepackage{multirow}
\usepackage{textcomp}
\usepackage{xcolor}
\usepackage[normalem]{ulem}

\colorlet{RevDelHL}{red!40}
\colorlet{RevAddHL}{green!35}
\colorlet{RevDelLine}{red!75!black}
\newcommand{\revdel}[1]{%
	\bgroup
	\markoverwith{%
		\hbox{%
			\color{RevDelHL}\rule[-0.45ex]{0.55em}{2.35ex}%
			\hspace{-0.55em}%
			\color{RevDelLine}\rule[0.55ex]{0.55em}{0.45pt}%
		}%
	}%
	\ULon{#1}%
}
\newcommand{\revadd}[1]{%
	\bgroup
	\markoverwith{\color{RevAddHL}\rule[-0.45ex]{0.55em}{2.35ex}}%
	\ULon{#1}%
}

\begin{document}

\title{LaminoDiff: Generative Computed Laminography via Near-Isotropic Spectral Supervision and Anisotropic Geometry}

\author{Tan~Liu,
        Liu~Shi,
        Binghuang~Peng,
        Tong~Jia,
        Xiaoling~Xu,
        Baodong~Liu,
        and~Qiegen~Liu%
\thanks{This work was supported by the National Natural Science Foundation of China (Grant nos.~621220033 and 62201193), the Nanchang University Youth Talent Training Innovation Fund Project (Grant no.~XX202506030012), and the Early-Stage Young Scientific and Technological Talent Training Foundation of Jiangxi Province (Grant no.~20252BEJ730005). Data were also provided by the Institute of High Energy Physics, Chinese Academy of Sciences, Beijing, China.}%
\thanks{T. Liu, L. Shi, B. Peng, X. Xu, and Q. Liu are with the School of Information Engineering, Nanchang University, Nanchang 330031, China (e-mail: liutan@email.ncu.edu.cn; shiliu@ncu.edu.cn; bh\_peng@email.ncu.edu.cn; xuxiaoling@ncu.edu.cn; liuqiegen@ncu.edu.cn).}%
\thanks{T. Jia and B. Liu are with the Institute of High Energy Physics, Chinese Academy of Sciences, Beijing 100049, China, the School of Nuclear Science and Technology, University of Chinese Academy of Sciences, Beijing 100049, China, and the Jinan Laboratory of Applied Nuclear Science, Jinan 250100, China (e-mail: jiatong@ihep.ac.cn; liubd@ihep.ac.cn).}%
\thanks{T. Liu and L. Shi are co-first authors.}%
\thanks{Corresponding authors: Liu Shi and Qiegen Liu.}%
}

\maketitle

\begin{abstract}
Computed Laminography (CL) is widely used for nondestructive inspection of extended planar objects, but its restricted angular coverage produces an anisotropic point spread function, missing-cone spectral incompleteness, and severe aliasing with interlayer leakage. This paper presents LaminoDiff, a physics-constrained diffusion framework for CL reconstruction. During training, a CT-derived near-isotropic supervision target is reconstructed from full-angle Computed Tomography (CT) projections physically degraded to match CL detector noise and focal-spot blur while retaining full angular coverage; it is withheld from inference. At inference, the reverse process uses only the CL observation. An Anisotropic Representation (AR) constructs depth-aware channels from three adjacent slices: a neighbor average, a center-neighbor residual, and the retained current slice, followed by directional in-plane feature extraction. Experiments on simulated and real multilayer printed circuit board data, including ball grid array and high-frequency stub samples, show that LaminoDiff improves artifact suppression, edge preservation, and depth stratification over analytic Feldkamp--Davis--Kress and representative learning-based baselines.
\end{abstract}


\section{Introduction}
\begin{figure}
	\centering
	\includegraphics[width=\linewidth]{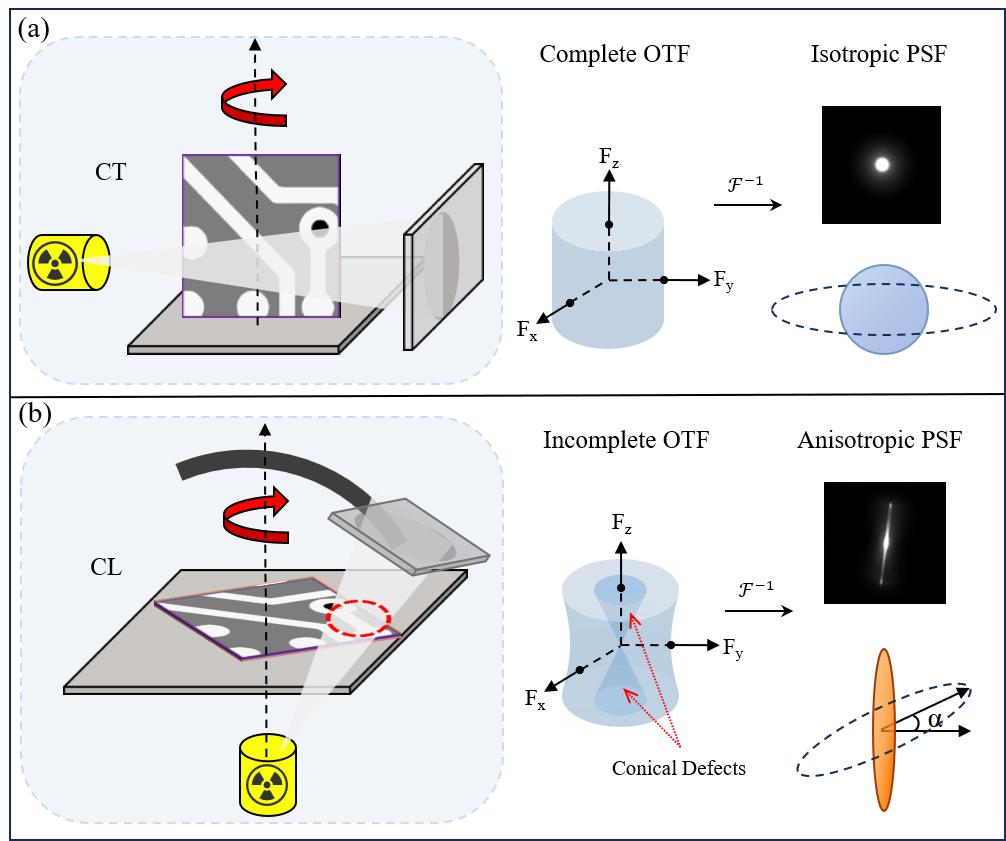}
\caption{%
		Comparison of scanning geometries and physical degradation between (a) CT and (b) CL.
		In contrast to the broader angular coverage of CT, which yields a more complete spectral response and a less anisotropic PSF, the inclined scanning geometry of CL results in spectral incompleteness (conical defects) in the Fourier domain. This geometric limitation fundamentally manifests as an elongated, anisotropic PSF in the spatial domain, leading to aliasing artifacts.
	}
	\label{fig:problem}
\end{figure}

\IEEEPARstart{C}{omputed} Laminography (CL) is used for nondestructive inspection of extended planar components, including high-frequency printed circuit boards and semiconductor wafers \cite{zhou1996computed,aryan2018overview,cai2020recent,du2025x}. In multilayer printed circuit board (PCB) inspection, elongation of the depth-wise point spread function (PSF) can produce interlayer leakage that obscures layer assignment, the continuity of vias, and the termination planes of high-frequency stubs. As illustrated in Fig.~\ref{fig:problem}, Computed Tomography (CT) \cite{kak2001principles} provides broader angular coverage and therefore a more complete spectral response and a less anisotropic PSF, whereas CL employs an inclined geometry. This configuration reduces the excessive X-ray attenuation paths that can occur in planar objects. Image reconstruction nevertheless remains an ill-posed inverse problem because of the missing cone and incomplete satisfaction of the Tuy--Smith data-completeness conditions \cite{tuy1983inversion,smith2007image}. As depicted in the OTF comparison in Fig.~\ref{fig:problem}, these conical defects appear as spatial anisotropy in the PSF \cite{acciavatti2019non}. The resulting impulse response is elongated along the depth axis and can generate aliasing artifacts. The broader angular coverage of CT provides a more complete and less anisotropic spectral response and can therefore support a physically meaningful near-isotropic supervision target.

While the analytical Feldkamp--Davis--Kress (FDK) algorithm \cite{feldkamp1984practical} is widely adopted for its efficiency \cite{jia2020gpu}, it is susceptible to severe data incompleteness \cite{pengxiang2025study}. Although various improvements, such as truncation correction \cite{wang2025truncated}, interlayer deblurring \cite{zhao2018edge,zhao2021iterative}, and virtual reprojection \cite{sun2021reconstruction}, have been proposed to mitigate these issues, aliasing artifacts remain intrinsic to the reconstruction. To further suppress noise and artifacts, iterative algorithms based on variational principles have become mainstream \cite{qi2015sparse}. These approaches attempt to design regularizers tailored to CL imaging features, exploring strategies including structure tensors \cite{lu2023anisotropic,lu2023cone}, gradient sparsity \cite{xi2024pig,tan2024orthogonal}, decoupled axis smoothing \cite{liu20253}, and specific solvers for local scanning tasks \cite{wang2025conjugate}. Despite their theoretical effectiveness, recovering fine structural details from the ill-posed integral operator often requires substantial computational cost and delicate hyperparameter tuning, which may limit industrial scalability. Explicit CT--CL fusion can compensate for spectral incompleteness, but it requires jointly acquired and registered dual-modal projections. Such paired measurements are often unavailable in industrial CL inspection.

In recent years, deep learning methods have been applied to tomographic imaging \cite{jin2017deep,chen2017low}, including neural networks for different stages of the inspection workflow \cite{jia2023correction,ronneberger2015u,shi2023automatic,shi2024pcb,zou2025artifact,feng2025multi}. However, directly applying these generalized methods to mitigate CL aliasing artifacts presents specific physical challenges. The optimization of deterministic regression models is often constrained by the physical scarcity of paired ground-truth labels in industrial settings. CL reconstructions themselves retain missing-cone aliasing and are therefore unsuitable as supervision targets. To address this, prior works have explored unsupervised domain adaptation to synthesize pseudo-labels \cite{zhu2017unpaired}. Although these adversarial approaches are effective in aligning textural distributions, ensuring the strict structural fidelity required for precision nondestructive testing remains an ongoing challenge, as unconstrained synthesis processes may occasionally introduce structural variations \cite{zhang2018task}. Beyond the choice of supervision, generic denoising networks are designed for spatially uniform degradation and therefore have limited capacity to separate depth-wise interlayer leakage from legitimate in-plane circuitry. Feeding neighboring slices as additional input does not remove this coupling, because the subsequent feature extraction still treats the stacked observations as isotropic measurements.

Recognizing the potential of diffusion models in addressing complex inverse imaging problems \cite{song2019generative,ho2020denoising,song2020denoising,liu2023dolce,liu2025pwd}, we investigate a physics-constrained generative framework termed LaminoDiff. In this framework, near-isotropic spectral supervision and Anisotropic Representation (AR) are coupled in a single generative process. Reconstructions from the inclined geometry retain missing-cone aliasing and therefore cannot serve as physically consistent supervision targets; instead, a near-isotropic supervision target is reconstructed from full-angle CT projections that are physically degraded to reproduce the CL detector noise and focal-spot blur while retaining full CT angular coverage, and this target is used only for training. Because this target cannot be acquired at test time without registered dual-modal projections, the reverse process is driven by the CL measurement itself. The elongated PSF couples neighboring slices while projecting in-plane directional footprints. AR reads the CL observation under this geometry: three adjacent slices form a neighbor-shared component and a center-neighbor residual, while the original current slice is retained, and spatial features are extracted along orthogonal in-plane axes. Supervised by the CT-derived near-isotropic target and conditioned on AR, the reverse process is anchored to a physically consistent baseline and is guided to compensate for directional blurring and spectral incompleteness.

The main contributions of this work are summarized as follows:
\begin{itemize}
\item  Near-Isotropic Spectral Supervision: We construct a near-isotropic supervision target from full-angle CT projections that are physically degraded to reproduce the CL detector noise and focal-spot blur while retaining full CT angular coverage, and use it to constrain the LaminoDiff generative process. The target is used only during training, so inference remains driven by CL measurements.
\item  Anisotropic Representation: We construct a depth-aware representation from three adjacent CL slices. The average of the two neighboring slices forms a neighbor-shared component, and subtracting this component from the current slice yields a center-neighbor residual. The original current slice is retained as an additional channel, and directional features are extracted along the two in-plane axes. This representation is designed to expose depth-correlated leakage and directional PSF effects.
\end{itemize}

The rest of this paper is organized as follows: Section~\ref{sec:related_work} briefly reviews relevant research. Section~\ref{sec:Methodology} presents the theoretical framework and the proposed method. Section~\ref{sec:Experiments} presents comparative experimental results. Finally, Section~\ref{sec:Conclusion} discusses the proposed method and provides concluding remarks.

\section{Related Work} \label{sec:related_work}
\subsection{CL Reconstruction and Physical Degradation}
CL reconstruction is inherently an ill-posed inverse problem. Given the object $x \in \mathbb{R}^N$ and measurements $y \in \mathbb{R}^M$, the forward process is modeled as:
\begin{equation}
	y = \mathcal{A}x + \eta_{\mathrm{meas}},
\end{equation}
where $\mathcal{A}$ is the CL system matrix and $\eta_{\mathrm{meas}}$ denotes measurement noise. Analytically, the FDK algorithm approximates the solution via filtered back projection:
\begin{equation}
	x_{fdk} = \mathcal{B} \left( \mathcal{H} * (\mathcal{W} \cdot y) \right),
\end{equation}
where $\mathcal{W}$, $\mathcal{H}$, and $\mathcal{B}$ represent the geometry correction weight, the filter, and the back projection operator, respectively.

The degradation mechanisms of $x_{fdk}$ can be quantitatively characterized by the system PSF. The physical mapping between the reconstructed image $x_{fdk}(\mathbf{r})$ and the original object $x(\mathbf{r})$ is modeled as a Fredholm integral equation of the first kind:
\begin{equation}
	x_{fdk}(\mathbf{r}) = \int_{\Omega} x(\mathbf{r}') h(\mathbf{r}, \mathbf{r}') d\mathbf{r}' + \eta_{\mathrm{recon}}(\mathbf{r}),
\end{equation}
where $\mathbf{r}$ represents the three-dimensional spatial coordinates, $h(\mathbf{r}, \mathbf{r}')$ is the system PSF, and $\eta_{\mathrm{recon}}(\mathbf{r})$ denotes reconstruction-domain noise and modeling error.

Restricted-angle acquisition elongates the impulse response along the depth axis and renders the PSF spatially non-stationary:
\begin{equation}
	h(\mathbf{r}, \mathbf{r}') \neq h(\mathbf{r} - \mathbf{r}').
\end{equation}

From an operator theory perspective, the blurring process induced by the PSF corresponds to a compact integral operator whose singular values $\sigma_k$ rapidly decay toward zero, and can be expressed as $\lim_{k \to \infty} \sigma_k = 0$ \cite{joyce2017pointspreadfunctionestimation}. This severe decay fundamentally renders the recovery of high-frequency structural details a highly ill-posed task. In the Fourier domain, the missing-cone effect provides a useful interpretation of the resulting directional spectral incompleteness. Because the PSF is spatially varying, the Fredholm model is not globally equivalent to a convolution and cannot be represented exactly by a single Fourier-domain multiplication. For a local spectral interpretation, we therefore use a local shift-invariant approximation within a small reconstruction region. Let $\mathbf{k}$ denote the spatial-frequency variable and $\theta_{\mathrm{tilt}}$ the CL tilt angle. Under this approximation, the effective relation is written as:
\begin{equation}
	\mathcal{F}(x_{fdk})(\mathbf{k};\mathbf{r}_0) \approx \mathcal{F}(x)(\mathbf{k};\mathbf{r}_0) \cdot \mathcal{M}_{\mathrm{eff}}(\mathbf{k};\mathbf{r}_0,\theta_{\mathrm{tilt}}) + \widehat{\delta}_{a}(\mathbf{k};\mathbf{r}_0),
\end{equation}
where $\mathcal{F}(\cdot)$ denotes the Fourier transform, $\mathcal{M}_{\mathrm{eff}}(\mathbf{k};\mathbf{r}_0,\theta_{\mathrm{tilt}})$ is an effective position- and geometry-dependent spectral response rather than an exact binary mask, and $\widehat{\delta}_{a}(\mathbf{k};\mathbf{r}_0)$ denotes the Fourier-domain direction-dependent aliasing residual caused by incomplete angular sampling. Consequently, the challenge of CL reconstruction is to recover spectral components weakly represented by $\mathcal{M}_{\mathrm{eff}}$ while mitigating the structured interference from $\widehat{\delta}_{a}$.

FDK reconstruction is widely used for its efficiency \cite{feldkamp1984practical,jia2020gpu}, while truncation correction, interlayer deblurring, and virtual reprojection mitigate related artifacts \cite{wang2025truncated,zhao2018edge,zhao2021iterative,sun2021reconstruction}. Iterative CL reconstruction uses SART or conjugate-gradient weighted least squares for data fidelity \cite{andersen1984simultaneous,wang2025conjugate}, with regularized variants imposing anisotropic total variation, directional penalties, or gradient sparsity \cite{lu2023anisotropic,lu2023cone,tang2024aawtv,liu20253,xi2024pig,tan2024orthogonal}. Prior-image and CAD-guided directional regularization have also been studied \cite{gui20233,xi2025cdtv}. CT--CL fusion can restore missing-cone content, but requires jointly acquired and registered projections \cite{jia2023multi,zuber2017augmented,ji2024fusional}. Because such dual-modal inference is often unavailable, we use CT-derived supervision with CL-only inference.

\subsection{Diffusion Models for Inverse Imaging}

Score-based generative models and denoising diffusion probabilistic models learn a prior over high-dimensional images by reversing a multi-scale noising process \cite{song2019generative,ho2020denoising}. In ill-posed inverse imaging this prior regularizes solutions that measurements do not uniquely determine. A deterministic convolutional regressor instead maps a single degraded input to a point estimate and is bounded by the spectral support of its paired targets \cite{jin2017deep,chen2017low}. Unpaired adversarial translators offer another generative route when such pairs are scarce; they can align textural statistics, yet unconstrained synthesis may still perturb topology that inspection must preserve \cite{zhu2017unpaired,zhang2018task}.
The discrete forward process is specified by the one-step Markov kernel:
\begin{equation}
	q(x_t \mid x_{t-1}) = \mathcal{N}\bigl(x_t; \sqrt{\alpha_t}\, x_{t-1}, (1-\alpha_t)\mathbf{I}\bigr),
\end{equation}
where $\alpha_t$ is the signal-retention coefficient at diffusion step $t$. The reverse process is a learned Gaussian transition:
\begin{equation}
	p_\theta(x_{t-1} \mid x_t) = \mathcal{N}\bigl(x_{t-1}; \mu_\theta(x_t, t), \sigma_t^2 \mathbf{I}\bigr),
\end{equation}
which is trained to invert the kernel above \cite{ho2020denoising}. For inverse imaging, Bayes' rule splits the posterior score into a generative prior and a measurement-consistency term:
\begin{equation}
	\nabla_x \log p(x \mid y) = \nabla_x \log p(x) + \nabla_x \log p(y \mid x).
\end{equation}
Nondestructive testing also benefits from reproducible reconstructions of the same volume. Denoising Diffusion Implicit Models (DDIM) provide a deterministic discretization of the reverse flow \cite{song2020denoising}, and classifier-free guidance mixes conditional and unconditional score estimates \cite{ho2022classifier}. In reconstruction this mixing can be used as a practical surrogate when the likelihood score $\nabla_x \log p(y \mid x)$ is not available in closed form. These two devices form the operational backbone of a diffusion reconstructor in this setting.

Diffusion priors have been applied to limited-angle CT to address missing angles along a circular trajectory \cite{liu2023dolce,liu2025pwd}. In addition to similar spectral holes, CL elongates the PSF along depth, causing interlayer leakage. Therefore, if a score estimator reads the CL observation under a spatially uniform degradation model, interlayer leakage remains mixed with in-plane circuitry. This difference indicates that CL reconstruction requires a representation that can read its depth-wise degradation.

\section{Methodology} \label{sec:Methodology}
\subsection{Near-Isotropic Spectral Supervision}
\subsubsection{Construction of the Near-Isotropic Supervision Target}

A physically consistent supervision target constrains the generative trajectory. Inclined-geometry CL reconstructions retain direction-dependent spectral incompleteness and are therefore excluded as labels. Near-isotropic spectral support is obtained from full-angle CT projections \cite{jia2023multi}, whose broader angular coverage yields a less anisotropic effective PSF. Here, $x_{\mathrm{iso}}$ denotes a CT-derived near-isotropic target, not an exactly isotropic reconstruction or the ideal physical object. It is used only as $x_0$ during training and is withheld from the reverse-process condition.

Full-angle CT projections are degraded in projection space to reproduce CL detector noise and focal-spot blur while retaining CT angular coverage. Let $y_{\mathrm{ct}}$ denote the degraded projections and $\mathcal{A}_{\mathrm{ct}}$ their system matrix. An analytic FDK reconstruction provides support in frequencies missing or attenuated in CL and initializes OS-SART refinement as $x_{\mathrm{iso}}^{(0)}$ \cite{feldkamp1984practical,andersen1984simultaneous}. For ordered subset $s$, let $y_s$ and $\mathcal{A}_s$ denote the corresponding measurements and system rows. With relaxation $\lambda$ and diagonal normalization $N$,
\begin{equation}
	x_{\mathrm{iso}}^{(k+1)} = x_{\mathrm{iso}}^{(k)} + \lambda N^{-1} \mathcal{A}_{s}^{\mathsf{T}} \bigl( y_{s} - \mathcal{A}_{s} x_{\mathrm{iso}}^{(k)} \bigr)\, .
\end{equation}
The algebraic updates reduce inconsistency with $y_{\mathrm{ct}}$ and stop after a finite number of steps. The resulting $x_{\mathrm{iso}}$ is the near-isotropic supervision target $x_0$.

\begin{figure}
	\centering
	\includegraphics[width=\linewidth]{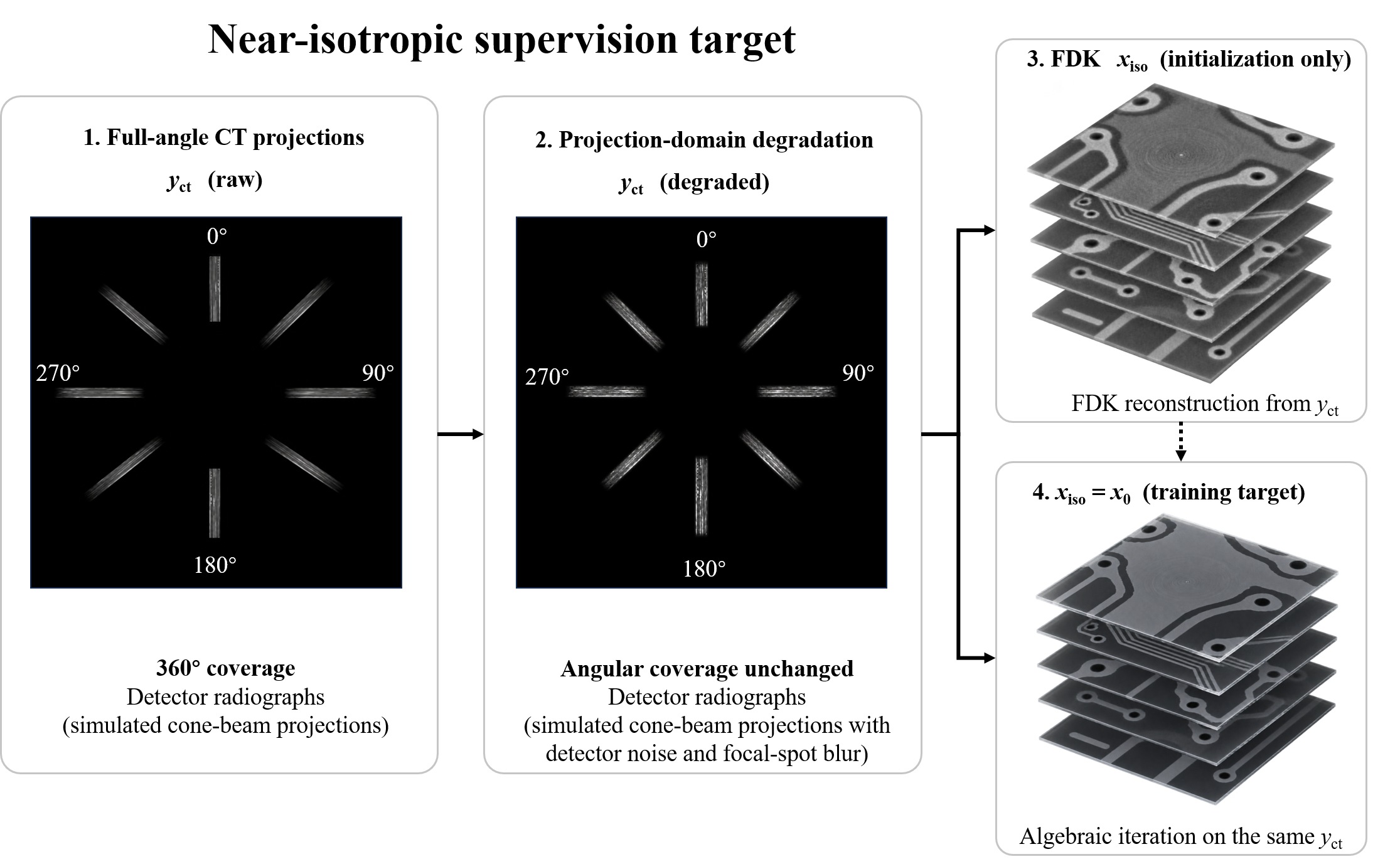}
	\caption{%
		Pipeline for constructing the near-isotropic supervision target.
		Full-angle CT projections undergo projection-domain physical degradation; the angular coverage is left unchanged.
		An FDK reconstruction of these projections provides support in the frequency region that is missing or strongly attenuated in CL and initializes algebraic iteration on the same measurements.
		The algebraic updates reduce residual inconsistency with $y_{\mathrm{ct}}$.
		The resulting volume $x_{\mathrm{iso}}$ is used only as the near-isotropic supervision target $x_0$ and does not enter the reverse-process condition.
	}
	\label{fig:iso_target}
\end{figure}

\subsubsection{Fidelity Analysis of Supervision Targets}

\begin{figure}
	\centering
	\includegraphics[width=\linewidth]{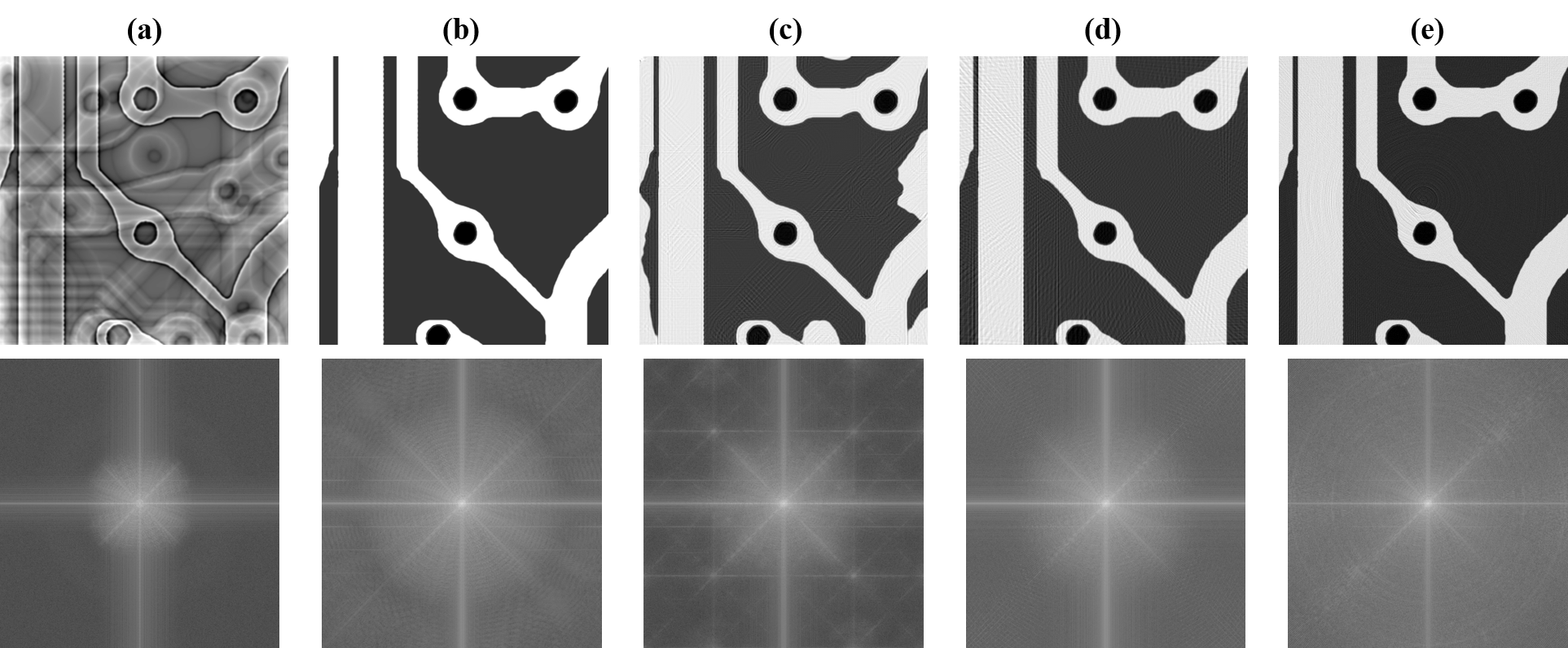}
	\caption{%
		Comparison of the CL observation and candidate supervision targets in the spatial (top) and frequency (bottom) domains.
		(a)~CL observation $x_{\mathrm{cl}}$ (analytic reconstruction of the inclined measurements): the input to be restored, not used as a label.
		(b)--(e)~Candidate supervision targets.
		(b)~Ideal binary reference volume $x_{\mathrm{bin}}$, which lacks measured X-ray statistics.
		(c)~Unpaired CycleGAN pseudo-label.
		(d)~FDK reconstruction of the physically degraded full-angle CT projections, used as $x_{\mathrm{iso}}^{(0)}$.
		(e)~Near-isotropic supervision target $x_{\mathrm{iso}}$, used as the supervision target $x_0$ in this work.
	}
	\label{fig:label_comparison}
\end{figure}
The reconstruction error of a trained network is bounded by how closely the network fits its supervision target and by how far that target sits from the ideal object \cite{zhang2017understanding,arpit2017closer,mohri2018foundations}. Let $f_\theta$ denote the neural network,  $x_{\mathrm{cl}}$ the artifact-contaminated CL observation, $x_{\mathrm{target}}$ the supervision target, and $x_{\mathrm{gt}}$ the ideal physical truth:
\begin{equation}
	\| f_\theta(x_{\mathrm{cl}}) - x_{\mathrm{gt}} \|_2 \leq \| f_\theta(x_{\mathrm{cl}}) - x_{\mathrm{target}} \|_2 + \| x_{\mathrm{target}} - x_{\mathrm{gt}} \|_2,
	\label{eq:error_bound}
\end{equation}
where the terms represent  the fitting error to the chosen supervision target and the bias of that target relative to the ideal object, respectively.
The bound separates fitting error from target bias \cite{lehtinen2018noise2noise}. It is tight only when the network fits $x_{\mathrm{target}}$ exactly. In Fig.~\ref{fig:label_comparison}(b)--(e), the second term serves as a geometric diagnostic. The binary reference sets this term to zero, but its piecewise-constant intensities lie outside measured X-ray statistics and increase the first term. The near-isotropic target $x_{\mathrm{iso}}$ is therefore used as $x_0$: it provides broader spectral support while remaining in the measurement domain. The ideal object $x_{\mathrm{gt}}$ is reserved for comparison and evaluation.

Fig.~\ref{fig:label_comparison} places the CL observation beside candidate supervision targets in the spatial and frequency domains. Panel~(a) shows $x_{\mathrm{cl}}$, the inclined-geometry reconstruction that the reverse process will read. Spatially it is low-contrast with interlayer leakage; its Fourier magnitude spectrum retains incomplete, orientation-dependent support relative to the full-angle CT-derived volumes. This panel is a reference for the input to be processed and is not treated as a supervision target. Supervision-target selection is among panels~(b)--(e). The binary reference volume $x_{\mathrm{bin}}$ in Fig.~\ref{fig:label_comparison}(b) provides ideal layouts with sharp piecewise-constant intensities outside measured X-ray statistics; its spectrum is broadband with strong axial structure from the hard edges. A network trained on it overfits unnatural boundaries. The CycleGAN pseudo-label in Fig.~\ref{fig:label_comparison}(c) can match local spatial contrast, while its Fourier spectrum exhibits unconstrained side lobes and its spatial field shows textural hallucinations; the translation is not tied to the projection geometry. The FDK volume in Fig.~\ref{fig:label_comparison}(d) is $x_{\mathrm{iso}}^{(0)}$: full-angle coverage already provides broader frequency support, while residual inconsistency with $y_{\mathrm{ct}}$ remains as spatial softness. The near-isotropic supervision target in Fig.~\ref{fig:label_comparison}(e) is obtained from the same degraded full-angle measurements after a finite number of algebraic updates from that FDK initialization. It retains the broader spectral support of panel~(d), sharpens circuit boundaries relative to panel~(d), and is used as the supervision target $x_{\mathrm{iso}}$.
\begin{figure*}[t]
	\centering
\includegraphics[width=0.85\linewidth]{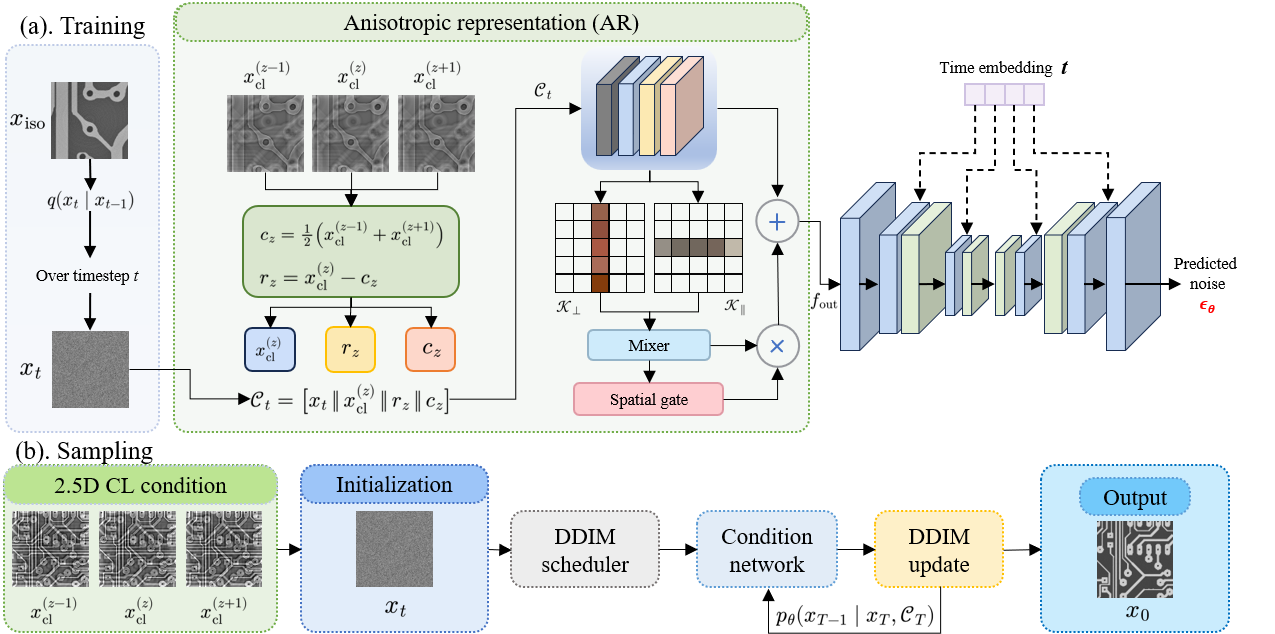}
	\caption{%
		Overview of the LaminoDiff framework.
		(a)~Training pipeline.
		Forward diffusion (left) perturbs the near-isotropic supervision target $x_0=x_{\mathrm{iso}}$ into $x_t$.
		Three adjacent CL slices supply a fixed observation-derived condition to AR (center): $\mathcal{C}_{\mathrm{obs}}$ contains the current slice, the neighbor-shared component $c_z$, and the center-neighbor residual $r_z$. At each reverse step, the current latent $x_t$ is concatenated with this fixed condition to form $\mathcal{C}_t$; the directional responses and spatial gate then produce the step-dependent representation $h_t$.
		AR runs at full resolution before a U-Net, conditioned on timestep embedding $t$, predicts $\epsilon_\theta$.
		(b)~Inference pipeline.
		The same observation-derived condition $\mathcal{C}_{\mathrm{obs}}$ is reused at every reverse step; the current latent changes from $x_t$ to $x_{t-1}$, so the AR representation $h_t$ is recomputed at each step. $x_T\sim\mathcal{N}(0,\mathbf{I})$ initializes the latent.
		Each reverse step applies the DDIM scheduler, the same AR condition network and U-Net readout, and a DDIM update until $x_0$; the near-isotropic supervision target is not supplied at inference time.
	}
	\label{fig:laminodiff_framework}
\end{figure*}

\subsection{Physically Constrained Score Estimation via AR}

The reverse process is trained to recover the near-isotropic supervision target $x_0=x_{\mathrm{iso}}$. The score estimator reads the CL observation through AR, using the same observation-derived condition during training and inference. As illustrated in Fig.~\ref{fig:laminodiff_framework}(a), the noisy state at timestep $t\in[1,T]$ is sampled by the reparameterization
\begin{equation}
	x_t = \sqrt{\bar{\alpha}_t} x_0 + \sqrt{1 - \bar{\alpha}_t} \epsilon, \quad \epsilon \sim \mathcal{N}(0, \mathbf{I}),
\end{equation}
where $\epsilon$ denotes standard Gaussian noise and $\bar{\alpha}_t$ is the cumulative signal-retention coefficient.

The elongated PSF couples neighboring slices and imprints orientation-specific aliasing. AR processes these effects at full slice resolution before U-Net downsampling (Fig.~\ref{fig:laminodiff_framework}(a)). Let $x_{\mathrm{cl}}^{(z-1)}$, $x_{\mathrm{cl}}^{(z)}$, and $x_{\mathrm{cl}}^{(z+1)}$ denote three consecutive slices. Their neighbor-shared component is
\begin{equation}
	c_z = \tfrac{1}{2}\bigl(x_{\mathrm{cl}}^{(z-1)}+x_{\mathrm{cl}}^{(z+1)}\bigr).
	\label{eq:common}
\end{equation}
Subtracting this component from the current slice defines the residual
\begin{equation}
	r_z = x_{\mathrm{cl}}^{(z)}-c_z.
	\label{eq:residual}
\end{equation}
Together, these equations define a depth-aware, rather than physically exact, layer split:
\begin{equation}
	x_{\mathrm{cl}}^{(z)} = r_z + c_z.
	\label{eq:slice_split}
\end{equation}
The channel $c_z$ captures neighbor-shared leakage-related structure, whereas $r_z$ captures current-slice-specific structure and local discrepancies; both may contain other depth-correlated content. Retaining the raw current slice, the fixed observation condition is
\begin{equation}
\mathcal{C}_{\mathrm{obs}} = \bigl[ x_{\mathrm{cl}}^{(z)} \,\Vert\, r_z \,\Vert\, c_z \bigr] \in\mathbb{R}^{3\times H\times W}.
\label{eq:obs_condition}
\end{equation}
At reverse step $t$, the current latent is concatenated with this condition:
\begin{equation}
\mathcal{C}_t = \bigl[ x_t \,\Vert\, \mathcal{C}_{\mathrm{obs}} \bigr] \in\mathbb{R}^{4\times H\times W}.
\label{eq:condition}
\end{equation}
Thus, $\mathcal{C}_{\mathrm{obs}}$ is fixed by CL, whereas $\mathcal{C}_t$ and $h_t$ vary with $x_t$; $\parallel$ denotes channel-wise concatenation.

AR reads $\mathcal{C}_t$ along both in-plane axes, whose kernel long axes match the slice coordinates; the tilt angle is fixed at run time.

Let $h_{\mathrm{in}}=\mathcal{C}_t\in\mathbb{R}^{4\times H\times W}$. Depthwise kernels elongated along the two in-plane axes produce
\begin{equation}
	\mathcal{D}_\parallel(h_{\mathrm{in}}) = \mathcal{K}_\parallel \circledast h_{\mathrm{in}}, \qquad
	\mathcal{D}_\perp(h_{\mathrm{in}}) = \mathcal{K}_\perp \circledast h_{\mathrm{in}},
	\label{eq:dir_resp}
\end{equation}
where $\circledast$ denotes depthwise cross correlation. Concatenation along the channel axis, followed by a mixer $\mathcal{P}$, yields
\begin{equation}
	\mathcal{D}(h_{\mathrm{in}}) = \mathcal{P} \bigl( \mathcal{D}_\parallel(h_{\mathrm{in}}) \oplus \mathcal{D}_\perp(h_{\mathrm{in}}) \bigr).
\end{equation}
The kernels are asymmetric along the two axes; $\oplus$ concatenates the responses and $\mathcal{P}(\cdot)$ fuses them into one directional correction.

A spatial gate $\mathcal{G}_\theta$ assigns a weight in $[0,1]$ to the directional response. The residual update is
\begin{equation}
	\begin{aligned}
		h_t
		&= \bigl(1-\mathcal{G}_\theta(\mathcal{D}(h_{\mathrm{in}}))\bigr)\odot h_{\mathrm{in}}\\
		&\quad + \mathcal{G}_\theta(\mathcal{D}(h_{\mathrm{in}}))\odot\bigl(h_{\mathrm{in}}+\mathcal{D}(h_{\mathrm{in}})\bigr)\\
		&= h_{\mathrm{in}} + \mathcal{G}_\theta(\mathcal{D}(h_{\mathrm{in}}))\odot\mathcal{D}(h_{\mathrm{in}})\, ,
	\end{aligned}
	\label{eq:dao_gating}
\end{equation}
where $\odot$ is the Hadamard product. A zero gate leaves $h_{\mathrm{in}}$ unchanged, whereas a unit gate applies $\mathcal{D}(h_{\mathrm{in}})$ fully. These operations constitute AR; $\mathcal{R}$ maps the latent and fixed observation condition to $h_t$.

A U-Net backbone estimates the score from $h_t$; timestep embeddings condition its encoder--decoder on noise level, coarse-resolution self-attention aggregates in-plane context, and skip connections preserve high-frequency layout.

Training matches the noise prediction to the near-isotropic supervision target. With $h_t\leftarrow\mathcal{R}(\cdot)$ as in Algorithm~\ref{alg:main},
\begin{equation}
	\mathcal{L}_{\epsilon} = \mathbb{E}_{t, x_0, \epsilon} \left[ \left\| \epsilon - \epsilon_\theta(h_t, t) \right\|_2^2 \right],
\end{equation}
where $\epsilon$ is standard Gaussian noise. This prediction implies a clean-image estimate
\begin{equation}
	\hat{x}_{0|t} = \frac{1}{\sqrt{\bar{\alpha}_t}}\Bigl(x_t-\sqrt{1-\bar{\alpha}_t}\,\epsilon_\theta(h_t,t)\Bigr).
	\label{eq:x0hat}
\end{equation}
The boundary term $\mathcal{L}_{\mathrm{b}}$ places an edge-weighted penalty on high-gradient loci of this estimate,
\begin{equation}
	\mathcal{L}_{\mathrm{b}} = \mathbb{E}_{t,x_0,\epsilon}\Bigl[\bigl\| w(x_0)\odot\bigl(\hat{x}_{0|t}-x_0\bigr)\bigr\|_1\Bigr],
	\label{eq:Lb}
\end{equation}
where $w(x_0)$ increases with the in-plane edge strength of the near-isotropic supervision target. The total objective is $\mathcal{L}=\mathcal{L}_{\epsilon}+\lambda_{\mathrm{b}}\mathcal{L}_{\mathrm{b}}$.
Conditioning on $\mathcal{C}_t$ ties the reverse trajectory to the CL observation while allowing $h_t$ to vary with $x_t$; $x_0$ appears only in the forward perturbation.

\subsection{Iterative Reconstruction via Deterministic Sampling}

During inference, $\mathcal{C}_{\mathrm{obs}}$ remains fixed while $x_t$ changes and produces a new $h_t$ at each step. Starting from $x_T\sim\mathcal{N}(0,\mathbf{I})$, DDIM supplies $(\bar{\alpha}_t,\bar{\alpha}_{t-1})$, AR forms $h_t=\mathcal{R}(x_t,\mathcal{C}_{\mathrm{obs}})$, and the U-Net predicts the noise for the deterministic update. The near-isotropic target is not supplied at inference. This is deterministic DDIM sampling \cite{song2020denoising}.

We interpret the trained noise prediction network $\epsilon_\theta$ as a conditional score function estimator.
With $h_t=\mathcal{R}(x_t,\mathcal{C}_{\mathrm{obs}})$, the gradient of the log density conditioned on the observation-derived condition is approximated as
\begin{equation}
	\nabla_{x_t} \log p_t(x_t \mid \mathcal{C}_{\mathrm{obs}}) \approx -\frac{1}{\sqrt{1 - \bar{\alpha}_t}} \epsilon_\theta(h_t, t),\label{eq:score_definition}
\end{equation}
where $\bar{\alpha}_t$ denotes the cumulative signal-retention coefficient. The observation-derived condition $\mathcal{C}_{\mathrm{obs}}$ is fixed at every reverse step, while $h_t$ changes with $x_t$.

Classifier-free guidance \cite{ho2022classifier} combines conditional and unconditional network evaluations. Define
\begin{equation}
	h_t^{+}=\mathcal{R}(x_t,\mathcal{C}_{\mathrm{obs}}),\qquad h_t^{-}=\mathcal{R}(x_t,\mathbf{0}),
\end{equation}
where $\mathbf{0}$ denotes a zero observation condition. Inference combines the corresponding noise predictions as
\begin{equation}
	\hat{\epsilon}_t = \epsilon_\theta(h_t^{-}, t) + s \cdot \left[ \epsilon_\theta(h_t^{+}, t) - \epsilon_\theta(h_t^{-}, t) \right],\label{eq:rectified_score}
\end{equation}
Here $s\geq 1$ is the guidance scale. The difference $\epsilon_\theta(h_t^{+},t)-\epsilon_\theta(h_t^{-},t)$ is the increment associated with the observation-derived condition; $s$ amplifies that increment in the reverse update.

\begin{algorithm}[!t]
	\caption{LaminoDiff: Representation Learning and Iterative Reconstruction}
	\label{alg:main}
	\begin{algorithmic}[1]
		\renewcommand{\algorithmicrequire}{\textbf{Input:}}
		\renewcommand{\algorithmicensure}{\textbf{Output:}}
		\item[\textbf{Representation Learning}]
		\REQUIRE Paired dataset $\mathcal{D} = \{(x_{\mathrm{iso}}^{(i)}, x_{\mathrm{cl}}^{(i)})\}_{i=1}^{N}$.
		\ENSURE Optimized noise prediction network $\epsilon_\theta$.
		\REPEAT
		\STATE Sample $(x_{\mathrm{iso}}^{(i)}, x_{\mathrm{cl}}^{(i)})$ and timestep $t \sim \mathcal{U}(1, T)$.
		\STATE Sample noise $\epsilon \sim \mathcal{N}(0, \mathbf{I})$.
		\STATE $x_0 \leftarrow x_{\mathrm{iso}}^{(i)}$,\quad $x_t \leftarrow \sqrt{\bar{\alpha}_t} x_0 + \sqrt{1-\bar{\alpha}_t} \epsilon$
		\STATE $\mathcal{C}_{\mathrm{obs}} \leftarrow [x_{\mathrm{cl}}^{(i,z)}\,\Vert\,r_z^{(i)}\,\Vert\,c_z^{(i)}]$
		\STATE $h_t \leftarrow \mathcal{R}(x_t,\mathcal{C}_{\mathrm{obs}})$
		\STATE $\hat{x}_{0|t} \leftarrow (x_t - \sqrt{1-\bar{\alpha}_t}\,\epsilon_\theta(h_t, t)) / \sqrt{\bar{\alpha}_t}$
		\STATE $\mathcal{L} \leftarrow \mathcal{L}_{\epsilon}+\lambda_{\mathrm{b}}\mathcal{L}_{\mathrm{b}}$
		\STATE Update parameters $\theta$ via gradient descent on $\nabla_\theta \mathcal{L}$
		\UNTIL{converged}

	\item[\textbf{Iterative Reconstruction}]
	\REQUIRE CL observation $x_{\mathrm{cl}}$, trained network $\epsilon_\theta$.
	\ENSURE Reconstructed image $x_0$.
	\STATE $x_T \sim \mathcal{N}(0, \mathbf{I})$
	\FOR{$t = T, \dots, 1$}
	\STATE $\mathcal{C}_{\mathrm{obs}} \leftarrow [x_{\mathrm{cl}}^{(z)}\,\Vert\,r_z\,\Vert\,c_z]$
		\STATE $h_t \leftarrow \mathcal{R}(x_t,\mathcal{C}_{\mathrm{obs}})$
		\STATE $h_t^{+} \leftarrow h_t,\quad h_t^{-} \leftarrow \mathcal{R}(x_t,\mathbf{0})$
		\STATE $\hat{\epsilon}_t \leftarrow \epsilon_\theta(h_t^{-}, t) + s \cdot \bigl[ \epsilon_\theta(h_t^{+}, t) - \epsilon_\theta(h_t^{-}, t) \bigr]$
		\STATE $\hat{x}_{0|t} \leftarrow (x_t - \sqrt{1-\bar{\alpha}_t}\hat{\epsilon}_t) / \sqrt{\bar{\alpha}_t}$
		\STATE $x_{t-1} \leftarrow \sqrt{\bar{\alpha}_{t-1}} \hat{x}_{0|t} + \sqrt{1-\bar{\alpha}_{t-1}} \cdot \hat{\epsilon}_t$
		\ENDFOR
		\RETURN $x_0$
	\end{algorithmic}
\end{algorithm}

For the iterative generation, we employ the deterministic update rule utilizing the rectified noise $\hat{\epsilon}_t$. The latent state $x_{t-1}$ at the previous timestep is computed as follows:
\begin{equation}
	x_{t-1} = \sqrt{\bar{\alpha}_{t-1}} \left( \frac{x_t - \sqrt{1 - \bar{\alpha}_t} \cdot \hat{\epsilon}_t}{\sqrt{\bar{\alpha}_t}} \right) + \sqrt{1 - \bar{\alpha}_{t-1}} \cdot \hat{\epsilon}_t,
	\label{eq:ddim_update}
\end{equation}
where the first term represents the predicted clean image at the current step, explicitly constrained by $\hat{\epsilon}_t$ to assist in preserving high-frequency fidelity and spectral consistency. The second term defines the deterministic direction pointing to $x_t$.
The same $\mathcal{C}_{\mathrm{obs}}$ and $\mathcal{R}$ are reused at every reverse step, while $h_t$ is recomputed from the current latent.

The LaminoDiff workflow is summarized in Algorithm~\ref{alg:main}.

\section{Experiments} \label{sec:Experiments}
To evaluate the reconstruction performance of LaminoDiff, experiments are conducted on both simulated data and real acquisitions. Simulated experiments support quantitative comparison, whereas real experiments examine applicability under industrial CL imaging.

\begin{table}[t]
	\caption{Scanning parameters of the CL system.}
	\label{tbl1}
	\centering
	\renewcommand{\arraystretch}{1.2}
	\begin{tabular}{@{} l c @{}}
		\toprule
		Parameter & Value \\
		\midrule
		Source-to-rotation-center distance & 19.32~mm \\
		Focus-to-stage distance & 16.35~mm \\
		Focus-to-zero-plane distance & 201.63~mm \\
		Rotation radius & 10.32~mm \\
		Detector unit size & 0.0495~mm \\
		Detector pixel size & 0.005~mm \\
		Tilt angle & $30^\circ$ \\
		Number of projection views & 360 \\
		Exposure time per projection & $1~\mathrm{s}$ \\
		\bottomrule
	\end{tabular}
\end{table}

\begin{figure}[t]
	\centering
	\includegraphics[width=\columnwidth]{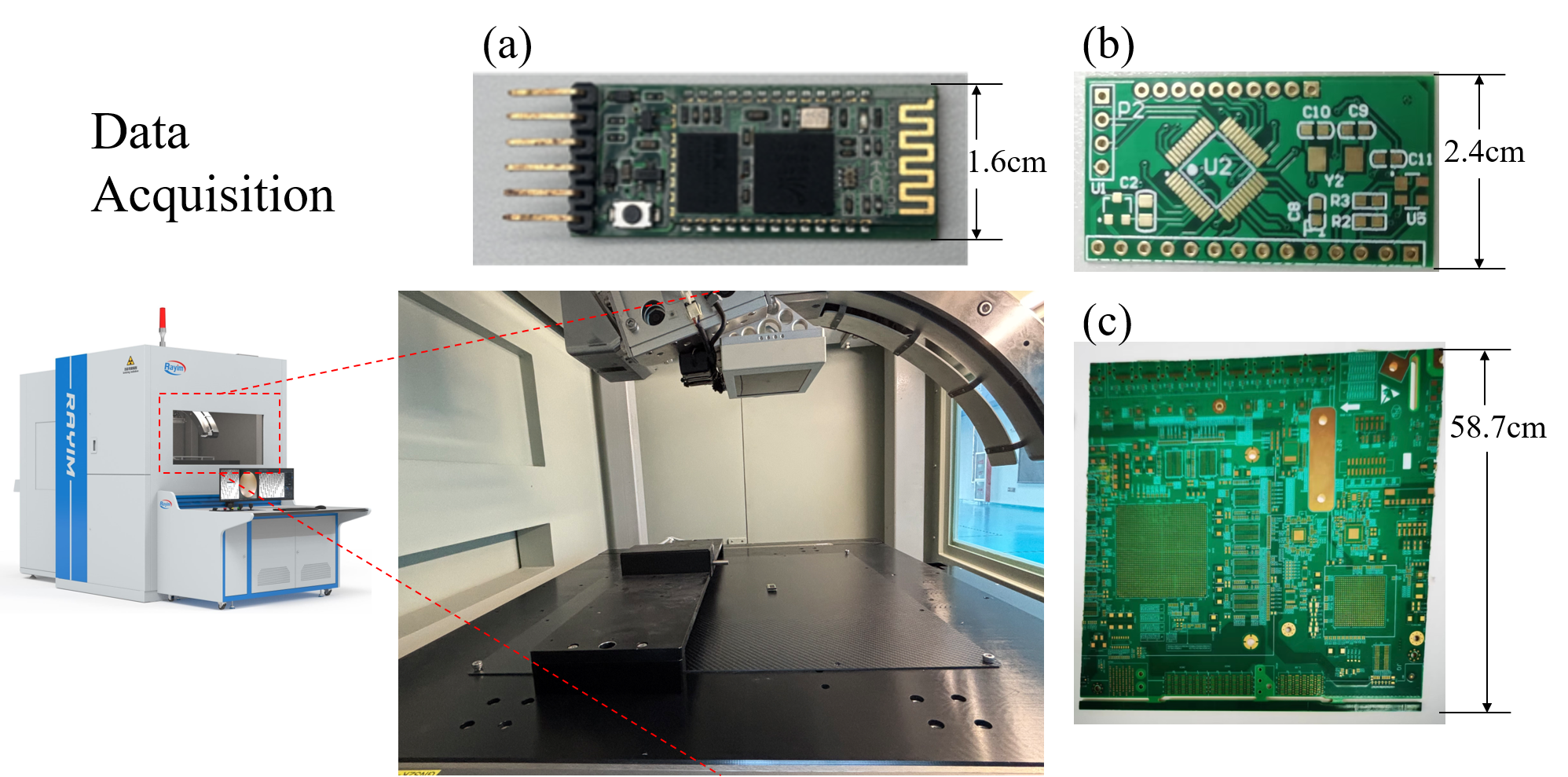}
	\caption{Optical photographs of the multilayer PCB samples used for real-world data acquisition.}
	\label{fig:cll_mach}
\end{figure}

\begin{figure*}[t]
	\centering
	\includegraphics[width=0.82\linewidth]{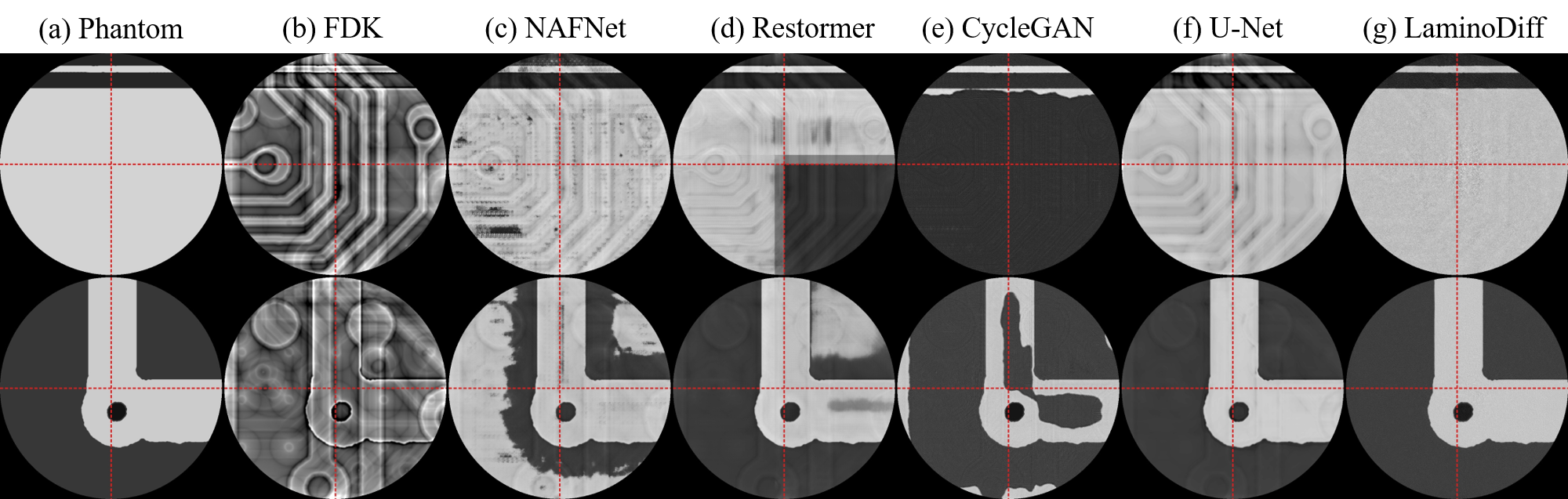}
	\caption{%
		Qualitative comparison on the homologous simulated reference volume.
		(a)~Reference volume;
		(b)~FDK;
		(c)~NAFNet;
		(d)~Restormer;
		(e)~CycleGAN;
		(f)~U-Net;
		(g)~LaminoDiff.
		The two rows correspond to two conductive layers.
	}
	\label{fig:ph20}
\end{figure*}

\begin{figure*}[t]
	\centering
	\includegraphics[width=0.82\linewidth]{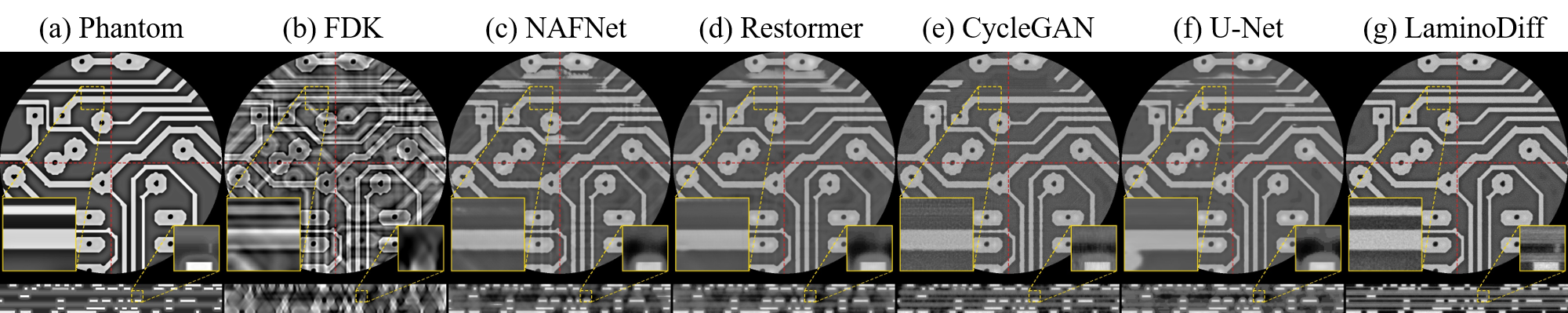}
	\caption{%
		Qualitative comparison on the heterogeneous simulated MCB reference volume.
		(a)~Reference volume;
		(b)~FDK;
		(c)~NAFNet;
		(d)~Restormer;
		(e)~CycleGAN;
		(f)~U-Net;
		(g)~LaminoDiff.
		The top row shows the in-plane reconstruction of a conductive layer, and the bottom row shows the vertical section.
	}
	\label{fig:mpcb}
\end{figure*}

\subsection{Experimental Setup and Evaluation Metrics}
LaminoDiff is implemented in PyTorch and trained and tested on a workstation equipped with one NVIDIA RTX 5880 Ada GPU with 48~GB memory. The contribution of each method is evaluated under a common train/test split and the same acquisition settings. The compared restoration baselines include NAFNet \cite{chen2022simple} and Restormer \cite{zamir2022restormer}. Supervised baselines use the paired synthetic corpus, whereas CycleGAN is trained with the corresponding unpaired domains under that split.

On simulated data, peak signal-to-noise ratio (PSNR), structural similarity (SSIM), mean squared error (MSE), and binary difference map agreement (BDM) are reported. SSIM is computed according to the standard structural-similarity formulation \cite{wang2004image}. Each metric is computed for every test slice and then averaged. PSNR and MSE therefore quantify intensity error from complementary per-slice logarithmic and squared-error views, SSIM summarizes local structural similarity, and BDM measures geometric agreement of the conductor layout. BDM first binarizes the reconstruction $I$ and the reference layout $I_{\mathrm{ref}}$ using the same Otsu threshold $\tau$ \cite{otsu1975threshold}:
\begin{equation}
\begin{aligned}
	B_X(i,j) &= \begin{cases}
		1, & \text{if } X(i,j) > \tau,\\
		0, & \text{otherwise,}
	\end{cases}
\qquad X \in \{I, I_{\mathrm{ref}}\},
\end{aligned}
\end{equation}
where $B_{\mathrm{rec}}=B_I$ and $B_{\mathrm{ref}}=B_{I_{\mathrm{ref}}}$. The pixel-wise agreement between the binary maps is then defined as
\begin{equation}
	\mathrm{BDM} = \left(1 - \frac{\sum_{i,j} |B_{\mathrm{rec}}(i, j) - B_{\mathrm{ref}}(i, j)|}{N \times M}\right) \times 100\%,
\end{equation}
where $B_{\mathrm{rec}}$ and $B_{\mathrm{ref}}$ are the binarized reconstruction and reference masks, and $N\times M$ is the image size. Higher BDM indicates better recovery of the circuit layout. On synthetic reference volumes, $I_{\mathrm{ref}}$ is the ground-truth conductor image and its binary mask is used as the reference; BDM is not computed for real scans without registered layout truth.

\subsection{Datasets and Scanning Protocol} \label{sec:Datasets}

\subsubsection{Simulated data}
Simulated experiments share the cone-beam geometry, voxel size, and projection-noise settings listed in Table~\ref{tbl1}. The homologous corpus models six-layer high-density interconnect (HDI) PCBs: a stochastic layout engine varies routing, vias, and layer stacking to diversify in-plane topology. For each reference volume, inclined CL projections and physically degraded full-angle CT projections are simulated to obtain paired observations $x_{\mathrm{cl}}$ and supervision targets $x_{\mathrm{iso}}$. The corpus contains 110 reference volumes, of which 100 are used for training and 10 are held out for testing. In addition, a heterogeneous multilayer complex board reference volume (MCB) is built from a complex PCB benchmark outside the homologous generator; its conductor density and routing complexity exceed the training distribution, while forward projection still follows Table~\ref{tbl1}. Homologous held-out reference volumes and the MCB reference volume form the simulated quantitative evaluation set.

\begin{table}[t]
	\caption{Quantitative evaluation on the homologous and heterogeneous simulated reference volumes.}
	\label{tab:quantitative_eval}
		\centering
		\renewcommand{\arraystretch}{1.2}
	\resizebox{\columnwidth}{!}{%
	\begin{tabular}{@{} l l c c c c @{}}
			\toprule
	Dataset & Method & PSNR $\uparrow$ & SSIM $\uparrow$ & MSE ($\times 10^{3}$) $\downarrow$ & BDM (\%) $\uparrow$ \\
			\midrule
			\multirow{6}{*}{Homologous} & FDK         & 13.18 & 0.8657 & 56.02 & 22.67 \\
			 & NAFNet      & 17.55 & 0.8352 & 50.59 & 81.39 \\
			 & Restormer   & 17.27 & 0.8744 & 59.97 & 93.76 \\
			 & CycleGAN    & 16.24 & 0.6715 & 84.64 & 82.46 \\
			 & U-Net       & 16.22  & 0.8573 & 65.42 & 99.71 \\
			 & LaminoDiff  & 27.04 & 0.8773 & 2.241 & 99.78 \\
			\midrule
			\multirow{6}{*}{Heterogeneous} & FDK         & 8.92  & 0.6769 & 128.3 & 83.86 \\
			 & NAFNet      & 17.85 & 0.8503 & 34.88 & 98.62 \\
			 & Restormer   & 17.12 & 0.8658 & 41.68 & 98.38 \\
			 & CycleGAN    & 18.26 & 0.7056 & 37.89 & 98.73 \\
			 & U-Net       & 15.98 & 0.8441 & 45.71 & 98.96 \\
			 & LaminoDiff  & 26.23 & 0.8277 & 1.363 & 99.27 \\
			\bottomrule
	\end{tabular}%
	}
\end{table}

\begin{figure*}[t]
	\centering
	\includegraphics[width=0.82\linewidth]{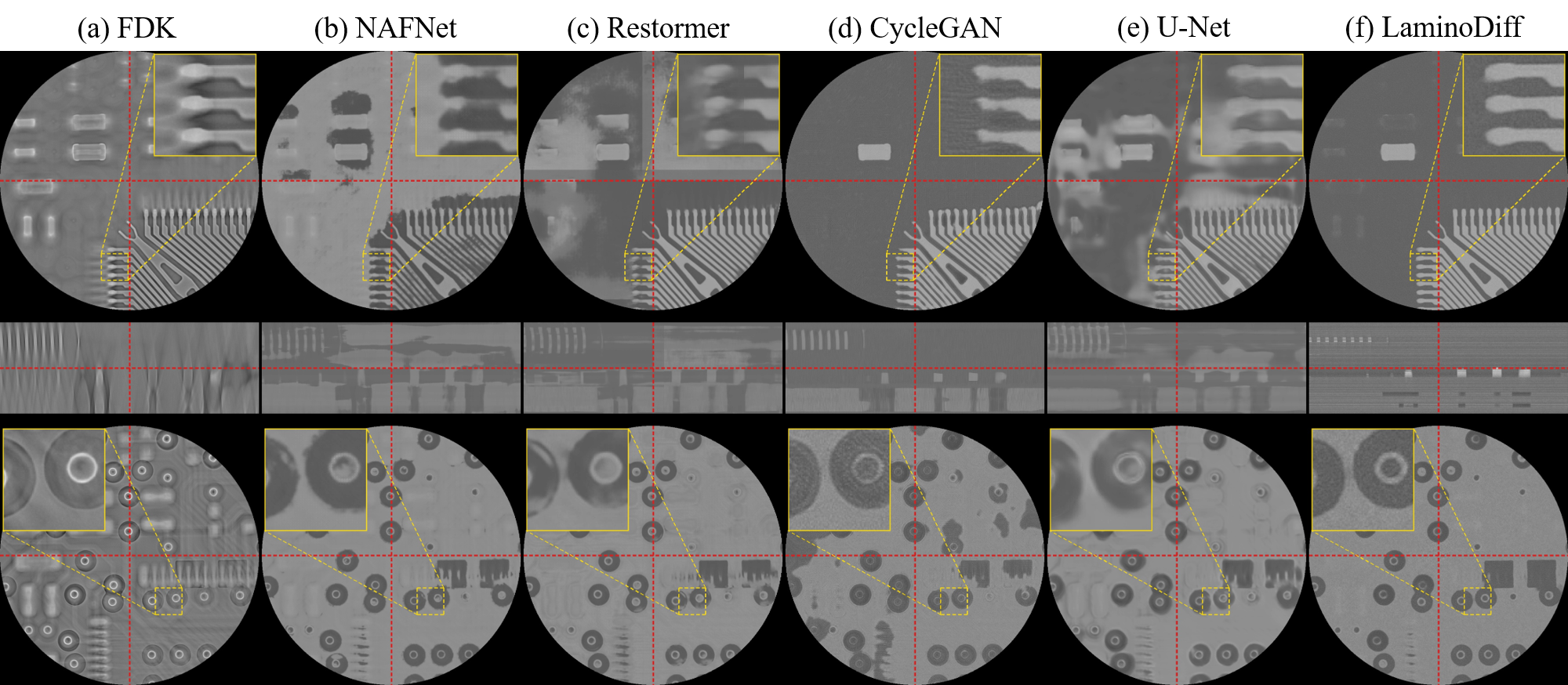}
	\caption{%
		Qualitative comparison on a real multilayer PCB.
		(a)~FDK;
		(b)~NAFNet;
		(c)~Restormer;
		(d)~CycleGAN;
		(e)~U-Net;
		(f)~LaminoDiff.
		The first and third rows show in-plane reconstructions of two conductive layers, and the second row shows the vertical section; yellow boxes mark magnified regions.
	}
	\label{fig:blue23}
\end{figure*}

\begin{figure*}[t]
	\centering
	\includegraphics[width=0.82\linewidth]{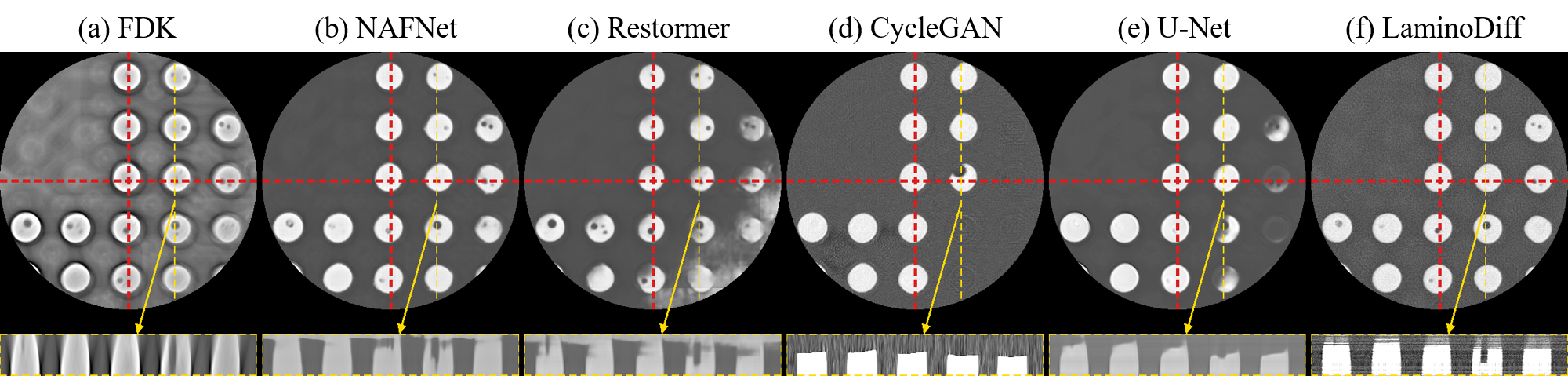}
	\caption{%
		Qualitative comparison on a real ball grid array (BGA) sample.
		Six columns from left to right: FDK, NAFNet, Restormer, CycleGAN, U-Net, and LaminoDiff.
		The first row shows the $xy$-plane reconstruction of the solder-ball array; the red dashed lines mark a common reference center, and the yellow vertical dashed line marks the $xz$ slice location.
		The second row shows the corresponding $xz$-plane vertical section extracted along the marked line.
	}
	\label{fig:bga}
\end{figure*}

\subsubsection{Real data}
Real experiments use three physical multilayer PCBs, as shown in Fig.~\ref{fig:cll_mach}, scanned at the Institute of High Energy Physics (IHEP), Chinese Academy of Sciences, Beijing. Data were acquired on a commercial industrial CL system at 80~kV tube voltage, 75~$\mu$A tube current, and a $30^\circ$ tilt angle; the source, rotation-center, stage, detector, and tilt settings follow the corresponding entries in Table~\ref{tbl1}. Real scans include focal-spot blur, electronic noise, and material heterogeneity not fully covered by simulation, and serve mainly for qualitative validation, with a semi-quantitative stub axial-length measurement.

\subsection{Simulated Experimental Results} \label{sec:SimResults}
This subsection qualitatively compares LaminoDiff with the baselines on simulated data. The reference volume denotes the ideal reference volume; homologous held-out volumes are in-distribution test samples, and the heterogeneous MCB volume is out-of-distribution.

\subsubsection{Homologous simulated reference volume}

Fig.~\ref{fig:ph20} compares two conductive layers on a homologous held-out volume. FDK shows orientation-specific aliasing, interlayer leakage, and low-contrast fine conductors. NAFNet and Restormer reduce some streaks but introduce noise or blocky distortion. CycleGAN improves contrast but produces structural breaks and textural hallucinations, whereas U-Net preserves the routing pattern with oversmoothed edges. LaminoDiff yields sharper boundaries, a cleaner background, and less leakage, consistent with Table~\ref{tab:quantitative_eval}.

\subsubsection{Heterogeneous simulated reference volume}
Fig.~\ref{fig:mpcb} shows the heterogeneous MCB volume, whose routing complexity exceeds the training distribution. FDK produces the strongest artifacts; NAFNet and Restormer soften vias and fine lines; CycleGAN introduces irregular textures; and U-Net loses edge detail. LaminoDiff better preserves narrow gaps and corners. In the vertical section, it separates multilayer interfaces more clearly and reduces leakage. The corresponding metrics are reported in Table~\ref{tab:quantitative_eval}.

Across Figs.~\ref{fig:ph20} and~\ref{fig:mpcb} and Table~\ref{tab:quantitative_eval}, LaminoDiff provides stronger artifact suppression, edge preservation, and interlayer separation than the baselines on both simulated datasets.

\begin{figure*}[t]
	\centering
	\includegraphics[width=\linewidth]{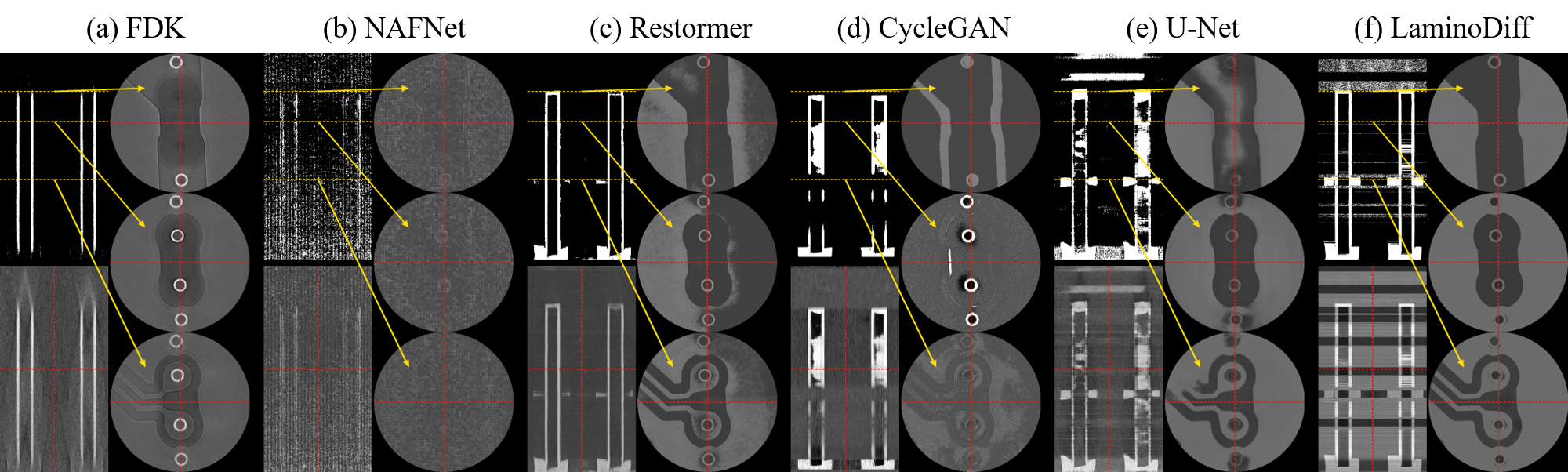}
	\caption{%
	Qualitative comparison on a real high-frequency stub structure.
	Six groups from left to right: FDK, NAFNet, Restormer, CycleGAN, U-Net, and LaminoDiff.
	Each group contains five panels: the left column contains the grayscale $xz$-plane vertical section (bottom), and the corresponding binary map (top) with yellow dashed lines marking three depths; the three circular panels on the right are $xy$-plane slices at those depths.
	}
	\label{fig:stub}
\end{figure*}

\begin{table*}[t]
	\caption{Ablation study on the heterogeneous MCB reference volume. CT-FDK denotes $x_{\mathrm{iso}}^{(0)}$, the analytic FDK reconstruction of physically degraded full-angle CT projections before OS-SART refinement; $x_{\mathrm{bin}}$ is the ideal binary reference in Fig.~\ref{fig:label_comparison}(b).}
	\label{tab:ablation}
	\centering
	\renewcommand{\arraystretch}{1.2}
	\begin{tabular}{@{} l l c c c c c @{} }
		\toprule
		Configuration & Target & AR & PSNR $\uparrow$ & SSIM $\uparrow$ & MSE ($\times 10^{3}$) $\downarrow$ & BDM (\%) $\uparrow$ \\
		\midrule
		FDK                         & --                       & -- & 8.92  & 0.6769 & 128.3 & 83.86 \\
		LaminoDiff ($x_{\mathrm{bin}}$) & $x_{\mathrm{bin}}$       & \checkmark & 20.44 & 0.9758 & 5.069 & 98.44 \\
		LaminoDiff ($x_{\mathrm{iso}}^{(0)}$) & $x_{\mathrm{iso}}^{(0)}$ & \checkmark & 22.39 & 0.7964 & 3.522 & 98.74 \\
		LaminoDiff (w/o AR)         & $x_{\mathrm{iso}}$       & $\times$ & 23.26 & 0.7393 & 3.450 & 98.75 \\
		LaminoDiff                  & $x_{\mathrm{iso}}$       & \checkmark & 26.23 & 0.8277 & 1.363 & 99.27 \\
		\bottomrule
	\end{tabular}
\end{table*}

\subsection{Real-World Experimental Results} \label{sec:RealWorld}
Real scans lack registered ground truth, so we compare them qualitatively and report a semi-quantitative stub axial-length measurement on the $xz$ section for multilayer PCB, BGA, and high-frequency stub samples.

\subsubsection{Multilayer circuit-board sections}
Fig.~\ref{fig:blue23} compares reconstructions on a real multilayer PCB. In-plane, FDK has low contrast, interlayer leakage, and orientation-specific aliasing; NAFNet and Restormer reduce streaks but blur fine lines; CycleGAN improves contrast but adds irregular textures; and U-Net recovers the main routing with oversmoothed adjacent lines. LaminoDiff gives clearer trace and pad boundaries, separates parallel fine lines in the magnified regions, and cleans the background. In the vertical section, it produces more compact layer stratification and weaker leakage than the baselines.

\subsubsection{BGA solder balls}
Fig.~\ref{fig:bga} compares reconstructions in a BGA region: the first row shows the $xy$-plane solder-ball array and the second the corresponding $xz$ section along the marked line. FDK shows divergent edges, ring artifacts, nonuniform background, and unresolved adjacent balls. NAFNet and Restormer recover more complete contours but retain soft edges and weak $xz$ contrast. CycleGAN raises contrast but adds irregular gaps and textures, while U-Net recovers ball shapes with oversmoothed boundaries and unclear internal voids. LaminoDiff produces a cleaner background, sharper $xy$ contours, more compact $xz$ transitions, and clearer voids, spacing, and fine-structure continuity.

\subsubsection{High-frequency stub structure}

Fig.~\ref{fig:stub} compares each method using the $xz$ vertical section, its binary map, and three depth-matched $xy$ slices. Under FDK, vertical stratification is blurred, the binary map contains fragmented contours and background speckles, and the $xy$ slices are nearly buried in artifacts; the start and end interfaces are indistinct, giving $L_z\approx 478.59\,\mu\mathrm{m}$. This manually calibrated FDK value is the reference because no registered CAD or ground-truth volume is available. NAFNet and Restormer provide limited noise reduction, leaving vertical interfaces, binary boundaries, and circular structures unclear. CycleGAN improves local contrast but interrupts vertical continuity and introduces irregular gaps and textural hallucinations. U-Net resolves the main horizontal layers and circular pads, but stub boundaries remain soft and leakage persists in the binary map. LaminoDiff more clearly delineates both stub interfaces, with more complete contours and a cleaner background, giving $L_z\approx 481.17\,\mu\mathrm{m}$, close to the FDK reference; its $xy$ slices also show clearer circular structures, connecting traces, and depth-wise differences.

Across Figs.~\ref{fig:blue23}--\ref{fig:stub}, on real industrial CL data without registered ground truth, LaminoDiff shows fewer visible artifacts and clearer in-plane detail, depth stratification, and structural continuity than the baselines. These observations support a qualitative assessment of multilayer PCB, BGA, and stub inspection.

\subsection{Ablation Study} \label{sec:Ablation}
An ablation study is conducted on the heterogeneous MCB reference volume to examine the roles of the supervision target and AR. Each variant changes only the corresponding component, while the remaining training and evaluation settings are kept identical; results are reported in Table~\ref{tab:ablation}. Analytic FDK is listed as a reference baseline.

For target selection with AR retained, the binary-reference variant preserves the structural layout well, but its piecewise-constant intensities provide limited gray-level fidelity because they do not follow the statistics of measured X-ray data. This indicates that structural supervision alone is insufficient for measurement-consistent reconstruction. Replacing the binary reference with the initial near-isotropic target improves the recovery of intensity values and reduces reconstruction error, but the remaining inconsistency in the target limits structural fidelity. The OS-SART-refined near-isotropic target provides a more suitable supervision signal: it retains broad spectral support while remaining closer to the measurement domain. Consequently, the full LaminoDiff model achieves the best overall balance among gray-level accuracy, structural consistency, and conductor-layout agreement, as summarized in Table~\ref{tab:ablation}.

With the refined near-isotropic target fixed, removing AR causes a clear degradation in reconstruction quality and increases the intensity error. The result indicates that the improvement of the full model does not arise from the supervision target alone. Instead, AR contributes complementary observation-derived information by exposing depth-correlated leakage and directional degradation across neighboring slices. Its inclusion therefore improves both the recovery of fine conductor structures and the separation of interlayer features. Overall, the ablation results show that the proposed method benefits from the combination of a measurement-consistent supervision target and an anisotropy-aware representation.

\section{Conclusion} \label{sec:Conclusion}
This work presents LaminoDiff, a generative framework for mitigating the ill-posed degradation in CL. The framework combines CT-derived near-isotropic spectral supervision with AR in a physically constrained generative process. AR reads depth-correlated components and in-plane directional footprints from the CL observation in a 2.5D manner, while the reverse process is guided toward the near-isotropic target under restricted-angle conditions.

Evaluations on simulated data and real industrial PCB measurements indicate stronger visible suppression of orientation-specific aliasing and interlayer leakage, with sharper conductor edges than analytic reconstruction and learning-based baselines. Simulated volumes support quantitative comparison; real PCB, BGA, and stub samples support qualitative assessment, with a semi-quantitative stub axial-length measurement.

The method has several limitations. Residual depth-wise anisotropy and interlayer crosstalk remain, and the missing cone still constrains vertical high-frequency recovery. AR uses only three adjacent slices, which limits long-range interlayer modeling. In addition, constructing the near-isotropic supervision target depends on paired full-angle CT data that industrial settings may lack. Future work will explore pseudo-3D constraints and interlayer attention to further suppress longitudinal crosstalk toward more isotropic three-dimensional resolution.

\bibliographystyle{IEEEtran}
\bibliography{references}

\end{document}